**Probing local lattice distortion in medium- and high-entropy alloys**

Y. Tong[1], G. Velisa[1], T. Yang[2], K. Jin[1], C. Lu[2], H. Bei[1], J. Y. P. Ko[3], D. C. Pagan[3], R. Huang[3], Y. Zhang[1,4], L. Wang[2], F. X. Zhang[1*]

[1]Division of Materials Science and Technology, Oak Ridge National Laboratory, Oak Ridge, TN 37831, USA.

[2]Department of Nuclear Engineering and Radiological Sciences, University of Michigan, Ann Arbor, MI 48109, USA.

[3]Cornell High Energy Synchrotron Source, Cornell University, Ithaca, NY 14850, USA.

[4]Department of Materials Science and Engineering, The University of Tennessee, Knoxville, TN 37996, USA.

*Corresponding to: zhangf@ornl.gov

**Abstract**

The atomic-level tunability that results from alloying multiple transition metals with d electrons in concentrated solid solution alloys (CSAs), including high-entropy alloys (HEAs), has produced remarkable properties for advanced energy applications, in particular, damage resistance in high-radiation environments. The key to understanding CSAs' radiation performance is quantitatively characterizing their complex local physical and chemical environments. In this study, the local structure of a FeCoNiCrPd HEA is quantitatively analyzed with X-ray total scattering and extended X-ray absorption fine structure methods. Compared to FeCoNiCr and FeCoNiCrMn, FeCoNiCrPd with a quasi-random alloy structure has a strong local lattice distortion, which effectively pins radiation-induced defects. Distinct from a relaxation behavior in FeCoNiCr and FeCoNiCrMn, ion irradiation further enhanced the local lattice distortion in FeCoNiCrPd due to a preference for forming Pd-Pd atomic pairs.

Local structure often deviates from average structure in complex materials with competing internal interactions, such as the competing electronic, magnetic and lattice interactions in the high-temperature superconducting materials[1]. These local deviations of structure change the nature of short range interactions and subsequently strongly influence the resultant properties of the material. Expanding our scientific knowledge of atomic-level heterogeneity from the relative simplicity of ideal, perfectly ordered, or structurally averaged materials to real materials should give rise to properties that meet the challenges at the frontiers of matter and energy[2-4]. When it comes to a new class of structural materials, compositionally complex HEAs, enthalpy among each atomic pair and size mismatch are not the only decisive factors to form single-phase concentrated solid solution alloys (SP-CSAs) but also short-range interactions to determine local chemical (i.e. atomic species in the nearest neighbor) and physical (i.e. bond distance) environments. HEAs owe their name to the large contribution of configurational entropy to the Gibbs free energy by mixing five or more elements to favor a single solid solution phase[5,6]. It is conceivable that compositional complexity can bring HEAs or other complex CSAs a relatively large fluctuation in the local chemical and physical environments, which offers a "plenty of room at the bottom"[7] to tune their properties by manipulating the complex local environment. Already, some CSAs have shown advanced and promising properties for extreme environment applications[8-13].

Local environment tuning has already benefited alloys' irradiation performance. For example, the adjustment of local chemical environment in concentrated alloys by mixing different elements or varying concentration dramatically increases their electrical resistivity and decreases their thermal conductivity[11,14], potentially assisting recombination of irradiation-induced point defects due to a prolonged thermal spike under irradiation[11]. Recently, FeCoNiCrMn (denoted by Mn-HEA hereafter) and FeCoNiCrPd (denoted by Pd-HEA hereafter) HEAs were reported to have the lowest electrical conductivity and thermal conductivity among the studied CSAs, including NiFe, NiCo, NiCoCr, NiCoFe and NiCoFeCr[14]. Consistently, Mn-HEA was found to exhibit the minimum swelling (< 0.2%) among other CSAs after $Ni^{2+}$ ion irradiation at 500℃[12,15]. On the other hand, undersized or oversized solute has been alloyed to austenitic or ferritic steels to suppress radiation-induced segregation at grain boundaries and void swelling by trapping point defects[4,16-20]. Moreover, large lattice distortions can retard dislocation movement to suppress the growth of damage structures[10,21]. Therefore, HEAs are expected to minimize damage evolution because their compositional complexity produces an intrinsically maximized lattice distortion. Strong local lattice distortion has been observed in body-centered cubic (BCC) ZrNbHf and NbMoTaW medium-entropy alloys (MEAs)[22,23], but it is not clear whether strong local lattice distortion exists in the face-centered cubic (FCC) HEAs because of limited data[24]. In the closely-packed FCC HEAs, the local distortion may be subtle because the enthalpy and size mismatch effects may interplay to relax the local strain, and a sophisticated quantitative measurement is thus required.

Rather than wandering the endless compositional landscape of CSAs aimlessly, we chose Pd-HEA as a starting point to study unexplored radiation behavior due to the hypothesized positive impacts from its large atomic size and low thermal conductivity. In the present study, we integrated two techniques, total scattering and extended X-ray absorption fine structure (EXAFS), to quantitatively characterize the local structure of Pd-HEA before and after ion irradiation. Also, the most-studied Mn-HEA and its base alloy, FeCoNiCr, were investigated for comparison. The elemental choice of Pd and Mn was intentionally studied to address the atomic size effect on local structure. In addition, a microstructural characterization with transmission electron microscope (TEM) was conducted on Mn- and Pd-HEAs after a prolonged ion irradiation at elevated temperature to evaluate the size mismatch impact on radiation defects.

**Results**

Pair distribution function (PDF) data obtained from a Fourier transformation of the total scattering data[1] provides a local view of the structure of these three compositional-complex CSAs in terms of interatomic distance (Fig. 1). Fig. 1a compares the observed PDFs with the ones calculated from a FCC structure with random arrangements of different elements, and their difference (green lines) shows some distinct features for these three alloys. The PDF difference for FeCoNiCr MEA has no visible r dependence, and random noise over the whole r range indicates a good fitting ($R_w$ ~7.3%) with the random structure model. Although the PDF difference of Mn-HEA shows a similar behavior as FeCoNiCr, a subtle difference exists at the short r range as revealed in later quantitative analyses. For Pd-HEA the difference in the PDF exhibits an apparent r-dependent behavior, especially a large difference in the first atomic shell, indicating a deviation of local structure from the average structure at long r range. An enlarged view of the local structure in Fig. 1b provides details of the local lattice distortion in Pd-HEA. The measured first PDF peak shifts to the high r direction with respect to the modeled peak, revealing a larger bond distance among the nearest atomic neighbors than expected in Pd-HEA.

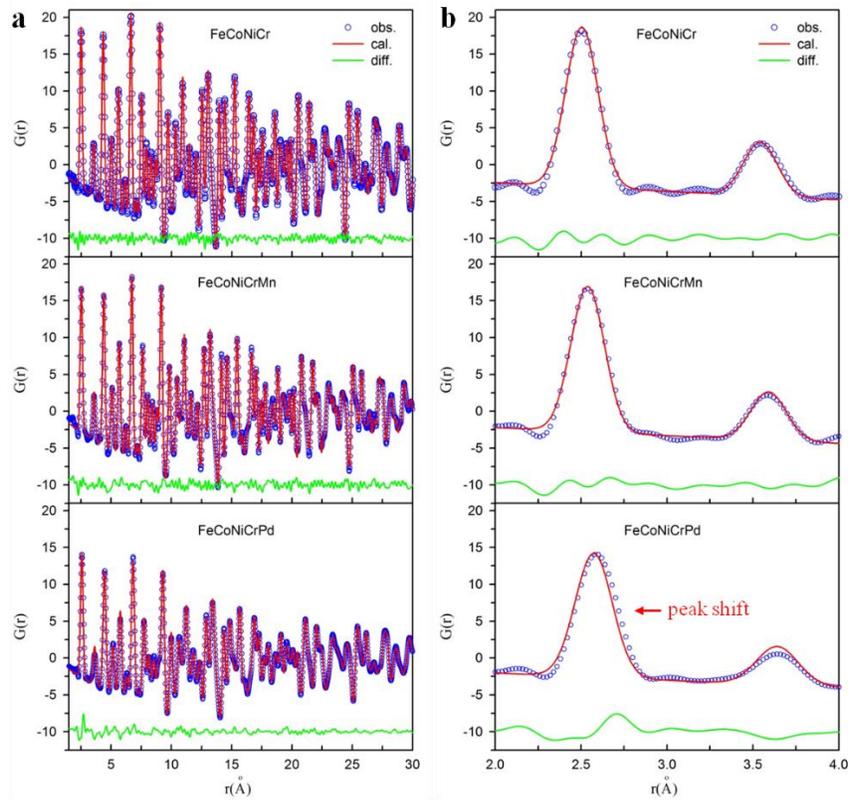

**Figure 1.** (a) A comparison between the observed X-ray PDF and the calculated PDF from a FCC structure model with identical occupancies of different atomic species. The Rietveld refinement ranges in r from 1.5 to 30 Å. (b) An enlarged view of the local structure in (a). The green line at the bottom of each figure shows the difference in PDF.

A commonly-used method to qualitatively analyze lattice distortion in materials is the measurement of the PDF peak width or the attenuation of the PDF peak intensity since they contain contributions from thermal and zero point displacements, as well as static displacement. The difficulty associated with this type of analysis is the separation of different displacement components[24]. Fig. 1 clearly shows that the addition of Pd into FeCoNiCr MEA strongly damps the envelop of PDF peaks while dramatically broadens the PDF peaks, and Mn has a very small effect (consistent with a recent neutron PDF measurement[24]). The Debye-Waller factor extracted from the Rietveld refinement ($\sigma^2_{\text{FeCoNiCr}}$ ~ 0.0065±0.0002 Å$^2$, $\sigma^2_{\text{Mn-HEA}}$ ~ 0.0066±0.0003 Å$^2$, and $\sigma^2_{\text{Pd-HEA}}$ ~ 0.0098±0.0003 Å$^2$) indicated that the $\sigma_{\text{Pd-HEA}}$ of Pd-HEA is larger by ~0.018 Å than those of FeCoNiCr and Mn-HEA. Commonly, heavy element leads to a small dynamic displacement, so we conclude that static displacement provides a major contribution to the peak broadening in Pd-HEA.

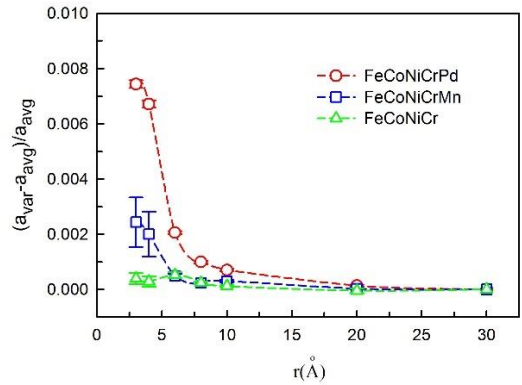

**Figure 2.** The r range in all Rietveld refinements starts from 1.5 Å but ends at different r value. $a_{\text{var}}$ denotes the r dependent lattice constant and $a_{\text{avg}}$ is the lattice constant of the average structure obtained from a full spectrum fitting.

The local distortion is also quantitatively analyzed by considering the differences in lattice constants extracted from varying r-range refinement, denoted by $a_{var}$. The major advantage of this quantitative method is the separation of static and dynamic displacements since the equilibrium position of atoms in a given crystal structure is only determined by the position of the PDF peaks. For the varying r-range refinement we fixed the $r_{min}$ to 1.5 Å but varied $r_{max}$ to gradually increase the weight of average structure. An $r_{min}$ of 1.5 Å was chosen to exclude the large noise at the low r range due to the Fourier transformation of the structure function with a finite Q range[1]. Additionally, the lattice constant of the average structure, $a_{avg}$, was obtained from a full spectrum fitting. The normalized variation of $a_{var}$ with reference to $a_{avg}$ is defined as lattice strain, $\varepsilon_{var} = (a_{var} - a_{avg})/a_{avg}$. The result in Fig. 2 shows the r dependence of $\varepsilon_{var}$. The FeCoNiCr MEA has a negligible lattice strain with the variation of fitting range. However, for Mn- and Pd-HEAs positive lattice strain shows up at short range, revealing that the local lattice is in tension, and this lattice distortion is strongly localized by a quick decay above the second atomic shell (~4 Å). In the Pd-HEA with a bond length of 2.6 Å, a 0.74% lattice strain in the 1st atomic shell should yield a static displacement of 0.019 Å, consistent with the rough estimation from the PDF peak width. Lastly, we calculated that the static displacement contributes ~1.2%, ~7.4% and ~20% to the $\sigma_{FeCoNiCr}$, $\sigma_{Mn-HEA}$ and $\sigma_{Pd-HEA}$, respectively.

For the Pd-HEA the local chemical environment surrounding Pd atoms was quantitatively analyzed by the Pd K edge EXAFS. It is a challenge to distinguish the Fe, Co, Ni and Cr atoms with EXAFS since their X-ray scattering factors are very close, so we simplified Pd-HEA as a pseudo-binary alloy $Pd_{20}X_{80}$ (X represents Fe, Co, Ni, and Cr) in the EXAFS spectrum fitting. Our fitting results (Table 1) show that the average number of Pd atoms neighboring with the excited Pd atom, $N_{Pd,1st}$ = 1.5, is very close to the random solid solution case, $N_{Pd,1st}$ = 1.6. Thus, the large lattice distortion in Pd-HEA should be a pure size mismatch effect.

**Table 1.** Number of Pd atoms neighboring the excited Pd atoms and bond length of different atomic pairs, and the average bond length among all atomic pairs in Pd-HEA were determined from EXAFS and PDF.

| Dose (dpa) | EXAFS | | | PDF |
| --- | --- | --- | --- | --- |
| | $N_{Pd,1st}$ | $R_{Pd-X}$ | $R_{Pd-Pd}$ | $R_{1st}$ |
| 0 | 1.5 | 2.60(5) | 2.70(9) | 2.597(1) |
| 0.1 | 2.0 | 2.60(9) | 2.70(8) | 2.606(4) |
| 0.3 | 2.1 | 2.60(1) | 2.69(3) | 2.609(6) |
| 1.0 | 2.1 | 2.59(7) | 2.69(1) | 2.614(3) |

The EXAFS spectrum was fit with a pseudo-binary model of $Pd_{20}X_{80}$.

An experimental study has demonstrated a substantial reduction of defect accumulation in NiCoCr MEA compared to pure Ni after irradiation with 1.5 MeV Ni ions at room temperature and further molecular dynamics (MD) simulations have revealed that the lattice distortion in NiCoCr MEA obstructs dislocation motion, leading to a slower growth of extended defects[10]. Due to a strong lattice distortion, Pd-HEA is expected to have an excellent capability to pin dislocations and slow down their growth. To verify this hypothesis, Mn- and Pd-HEAs were irradiated with 3 MeV $Ni^{2+}$ ions up to 55±5 dpa at the region of 700±100 nm from the sample surface. After such a prolonged irradiation at elevated temperature, dislocation loops in these two irradiated HEAs were examined by TEM, as shown in Fig. 3. Compared to Mn-HEA, Pd-HEA has a high density of dislocation loops (Pd-HEA: (19.6±2)×10$^{21}$ m$^{-3}$, and Mn-HEA: (5.83±0.6)×10$^{21}$ m$^{-3}$) but with a very small average size (Pd-HEA: 4.7±1 nm, and Mn-HEA: 15.9±1 nm). Besides, previous study showed that the average dislocation loop size in FeCoNiCr is twice larger than the one in Mn-HEA under 3 MeV Ni$^{2+}$ ions irradiation to 38±5 dpa at 773 K[10,21]. Large local lattice distortion in Pd-HEA can both directly and indirectly effect the growth of dislocation loops. Considering the direct pinning effect, the dislocation-local distortion interaction in Pd-HEA could be one order of magnitude higher than that in Mn-HEA because of a power-law relationship between solution strengthening ($\Delta\sigma_s$) and size mismatch ($\varepsilon_s$)[25], $\Delta\sigma_s \propto \varepsilon_s^{3/2}$. On the other hand, large local lattice distortion reduces the thermal conductivity of Pd-HEA by strongly scattering free electrons, which can prolong the thermal spike lifetime leading to an effective recombination of interstitials and vacancies at early defect evolution stage[11]. Furthermore, from the perspective of energy minimization oversized Pd atoms prefer trapping vacancies to release lattice strain, potentially increasing the probability of recombining vacancies with nearby interstitial loops. Note that vacancy diffusion can be accelerated by large atoms[26-29] and a balance between vacancy diffusion and dislocation pinning is needed to optimize the recombination probability. However, vacancy diffusion is difficult to be tuned due to joined influencing factors in alloys, e.g. electronic bonding, enthalpy and entropy, but dislocation pinning effect is much easier to be manipulated through a simple control of atomic size mismatch.

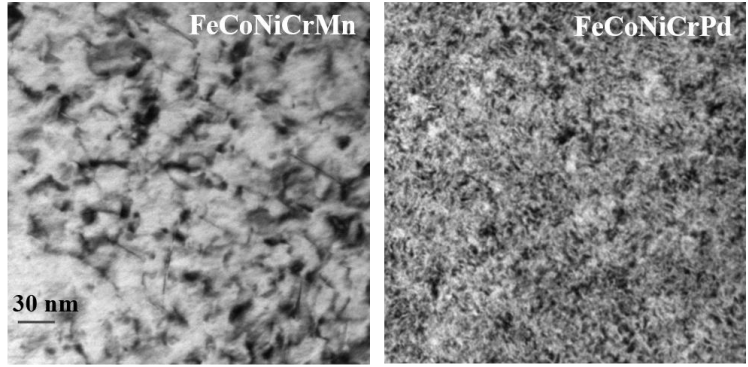

**Figure 3.** Dislocation loops in two-beam condition bright-field images of Mn- and Pd HEAs irradiated to 50±5 dpa at 700±100 nm from the surface at 693 K.

To further investigate radiation effects on the local structure of FeCoNiCr, Mn-HEA and Pd-HEA, irradiation with multiple energies of Ni ions (16 MeV, 8 MeV, 4 MeV and 2 MeV) to various fluences were conducted to produce a deep and uniform damage region (3.5 μm). The purpose of generating deep damage region is to enhance a signal-to-noise ratio in PDF and EXAFS measurements. PDF measurement shows that the lattice constant of the average structure increases with the radiation dose for all alloys, indicating the overall tensile strain is from irradiation-induced damage, in agreement with the MD simulations[30] and other X-ray diffraction measurements[31,32]. In Fig. 4, the plot shows the dose dependency of the local lattice strain in Pd-HEA. FeCoNiCr and Mn-HEA have a tendency to relax the local strain after irradiation while the local strain in Pd-HEA keeps increasing with the dose. We note that a local strain relaxation behavior was also observed in NiCoCr MEA (supplementary Fig. 1). Under prolonged irradiation, the buildup of local lattice strain in Pd-HEA may further improve its radiation tolerance by increasing the pinning capability. In addition, EXFAS results in Table 1 suggest that the increase of local lattice strain in Pd-HEA is related to the change of the local chemical environment of Pd atoms. The EXFAS fitting results show that number of Pd atoms neighboring excited Pd atoms, $N_{Pd,1st}$, increases up to 2.1 after irradiation while the bond distance of Pd-Pd pair and Pd-X pair exhibits a negligible change (Table 1). PDF results, however, show the average bond distance increases with dose, suggesting that a relatively large adjustment of the bond distance of the majority X-X pairs is required to accommodate the chemical environment change caused by large Pd atoms.

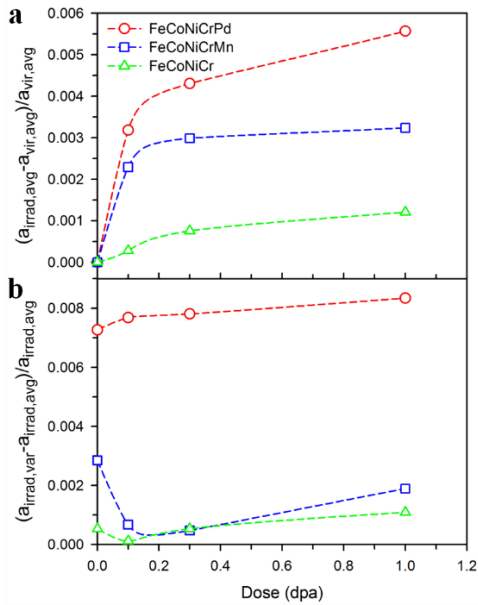

**Figure 4.** (a) Ion irradiation effect on the lattice constant of the average structure; (b) Ion irradiation effect on the lattice distortion in the 1st atomic shell. Here, $a_{irrad,var}$ was extracted from the Rietveld refinement in a r range from 1.5 Å to 3 Å, covering only the 1st atomic shell.

**Discussion**

Besides our quantitative measurement of lattice distortion, Zhang et al.[33] estimated the size mismatch effect in HEAs by a parameter

$$\delta\% = \sqrt{\sum_{i=1}^{N} c_i \left(1 - r_i / \sum_{j=1}^{N} c_i r_i \right)^2}$$

, where $N$ is the total number of the constituent elements, $c_i$ and $r_i$ denote the atomic fraction and atomic radius of the $i$th element, respectively. In addition, Ye et al.[34] calculated the fluctuation of local residual strain, $\varepsilon_{RMS}$, in HEAs from the perspective of packing efficiency. In Table 2, a comparison is made between our measurements and the calculations from the aforementioned two methods. A general agreement that local lattice distortion increases with increasing atomic size mismatch can be reached among the calculations and our measurement, but our experimental results are one order of magnitude smaller than their calculations. For Pd-HEA, $\delta\%$ and $\varepsilon_{RMS}$ of 4.46%[35] would lead to a static displacement of 0.116 Å, which is unreasonably larger than the experimental $\sigma_{Pd-HEA}$ (0.098 Å). The pronounced departure of the present measurements from the calculations may originate from missing charge transfer effect in these two models. These two models are based upon the hard-sphere concept without an electronic energy scale. Atomic size adopted in the calculations is empirically determined from some alloys or compounds, but atomic size of an element varies from one composition to another because of charge transfer, e.g., density functional theory calculations show that due to charge transfer Co atoms in FeCoNi MEA are under compressive strain while they are under tensile strain in FeCoNiCr MEA[36].

**Table 2.** A comparison of lattice distortion between predications and experimental results.

| Composition | Local distortion | | |
| --- | --- | --- | --- |
| | $\delta^{35}$ | $\varepsilon_{RMS}$ (%)[35] | $\varepsilon_{1st}$ (%) |
| FeCoNiCr | 0.3 | 0.39 | 0.04±0.02 |
| FeCoNiCrMn | 3.27 | 3.25 | 0.24±0.09 |
| FeCoNiCrPd | 4.46 | 4.46 | 0.74±0.01 |

1st denotes the 1st atomic shell.

Elastic field produced by an inclusion in a solid is elegantly described by Eshelby theory[37] within the framework of continuum elasticity. When examining an elastic field at atomic level, the validity of the continuum assumption should be verified since atoms are discrete particles. Through MD simulations some pioneering work has been conducted to extend the Eshelby theory to atomic level stresses in liquid without external stresses[38]. The MD simulation demonstrated that the atomic level stress in liquid follows Eshelby's theory but has additional component following an exponential decay with distance[38]. Here, we experimentally inspect the continuum assumption for the local strain field induced by relatively large size mismatch in Mn- and Pd-HEAs free from external perturbation. Fig. 5 compares the local strain field in these two HEAs with Elshelby's theory. Elshelby's inclusion theory proves that the displacement field away from an inclusion has a following relationship with respect to distance,

$$u(r) \propto 1/8\pi(1-\nu)r^2$$

, where $\nu$ is Poisson's ratio[37]. From Fig. 5, we can clearly see that the strain field caused by atomic size mismatch follows the law of $1/r^2$. Therefore, the local lattice distortion obeys the continuum mechanics.

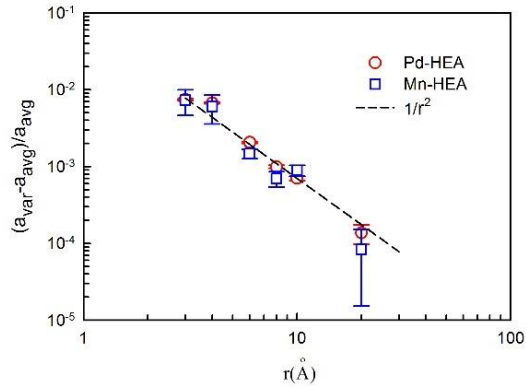

**Figure 5.** A comparison of the measured strain field and Eshelby theory. Plots are rescaled for data collapsing. Here, data for Mn- and Pd-HEAs are from Fig. 2.

A stable pinning effect is critical for a continuous recombination of radiation-induced defects in nuclear materials during their service time. Thus, a sharp decrease of local lattice distortion in FeCoNiCr, Mn-HEA and NiCoCr starting at a very low dose should be avoided. The reason for this reduction of pinning effect is the overall lattice expansion induced by radiation defects. One obvious solution is the design of alloys with a high enough local lattice distortion, which cannot be fully relaxed by lattice expansion. On the other hand, the change of the local chemical environment induced by radiation can generate positive effect on the pinning ability, including an aggregation of large-size atoms and a formation of short-range order (SRO). An irradiation-enhanced SRO was found in the NiCoCr CSA under the same ion-irradiation condition[39], but this SRO contribution to the pinning effect is not high enough to overcome the effect from the large lattice expansion (supplementary Fig. 1). A combination of the strong lattice distortion and the local chemical rearrangement, however, can be a good strategy to enhance the pinning effect, as suggested by the case of Pd-HEA.

To summarize, we have revealed the change of the local physical and chemical environments in a newly-developed Pd-HEA under irradiation based on a quantitative method combining PDF and EXAFS techniques. Large Pd atoms induce a strong local lattice distortion, 20% of the total atomic displacement including dynamic and static contributions. The local lattice distortion in Pd-HEA is further increased under ion irradiation due to a preference for forming Pd-Pd pairs. The local structure characterization has provided an atomic-level understanding of suppressed growth of extended defects in Pd-HEA. Results from TEM observation show a delay on the loop growth in Pd-HEA than Mn-HEA, suggesting the loops in Pd-HEA are in an earlier stage of loop evolution than a similarly implanted Mn-HEA sample. Predictively, the defect recombination in Pd-HEA can be further improved by optimizing the Pd concentration.

The present study provides new insights into radiation damage control by tuning the atomic level bond structure. Thus, understanding the critical roles of atomic-level heterogeneity beyond overly simplistic model materials is a pressing need. The ability to mediate radiation damage in compositionally complex alloy systems may lead to new research avenues, including science-based accelerated material discoveries and radical changes in the field of radiation effects.

## Methods

**Sample Preparation.** Elemental Ni, Co, Fe, Cr, Mn and Pd (>99% pure) were carefully weighted and mixed into NiCoFeCr, NiCoFeCrMn, and NiCoFeCrPd by arc melting. The arc-melted buttons were flipped and re-melted at least five times to improve homogeneity and then drop cast into a copper mold measuring 12.7×12.7×70 mm$^3$. The ingots were homogenized for 24 h at 1200 °C, and then rolled at room temperature in steps to a final thickness of ~1.8 mm. The rolled specimens were annealed at 800 °C for 1h. After above processing, all alloys show equiaxed grain structures with averaged grain sizes of approximately 6 µm, 8 µm and 2 µm for FeCoNiCr, FeCoNiCrMn and FeCoNiCrPd, respectively (supplementary Fig.2).

**Ion Irradiation.** Specimens after 3 MeV $Ni^{2+}$ ion irradiation up to 55±5 dpa at 693 K were used for the TEM microstructural characterization. And samples were irradiated with multiple energies of Ni ions (16 MeV $Ni^{5+}$, 8 MeV $Ni^{3+}$, 4 MeV $Ni^{1+}$ and 2 MeV $Ni^{1+}$) to various fluences (from 0.1 dpa to 1 dpa) at room temperature for total scattering and EXAFS studies. The details about multiple-energy irradiation can be found in refs 39,40. During the ion irradiation, a raster ion beam was used to ensure a homogeneous irradiation. All the irradiation experiments were carried out in the Ion Beam Materials Laboratory at the University of Tennessee. Predictions of local dose (dpa) in samples were calculated by the Stopping and Range of Ions in Matter 08 code (SRIM 08) in Kinchin-Pease option simulation mode[41] assuming a displacement threshold energy of 40 eV for all elements. The multiple-energy irradiation generated a region of approximately uniform damage to a depth of about 3500 nm based on SRIM damage profile calculation. The purpose of generating a deep damage region is to enhance a signal-to-noise ratio in total scattering and EXAFS measurements.

**Materials Characterization.** The total scattering measurement was carried out at F2 station of CHESS with an X-ray energy of 61.332 keV. A two-dimensional stationary detector with 200×200 $\mu m^2$ pixel size was used to collect data. Calibration was performed using $CeO_2$ NIST powder standard. Fit2D software[42] was used to correct for a beam polarization and a dark current. PDFgetX3[43] was used to obtain real-space atomic pair distribution function (PDF) by a Fourier transformation of the measured and then normalized reciprocal-space structure function in a Q range of 30 Å$^{-1}$. By using PDFGUI software[44] the measured PDFs were refined with Rietveld method. EXFAS measurement was conducted in a grazing fluorescence mode at F3 station of CHESS. The normalization of X-ray absorption spectra, extraction of EXAFS oscillations and data analysis were performed following standard procedures using Demeter software package[45]. Cross-sectional TEM foils were prepared by focused ion beam (FIB) lift-out techniques on a FEI Helios Nanolab Dualbeam workstation in the Michigan Center for Materials Characterization of the University of Michigan. Bright-field TEM imaging of dislocation loops were taken in a two-beam condition (**g** = [200]) using JEOL 3011. To minimize the artificial effects from surface sinks and injected interstitials[46], the region of 700±100 nm with a damage dose of 50±5 dpa-calculated by SRIM 08[41] was chosen for the statistic study of loop distribution.

**Acknowledgments**

This work was supported as part of the Energy Dissipation to Defect Evolution (EDDE), an Energy Frontier Research Center funded by the U.S. Department of Energy, Office of Science, Basic Energy of Sciences. The X-ray PDF and EXAFS measurements were conducted at the Cornell High Energy Synchrotron Source (CHESS) which is supported by the National Science Foundation and the National Institutes of Health/National Institute of General Medical Sciences under NSF award DMR-1332208. The microstructure characterization was conducted in the Michigan Center for Material Characterization of the University of Michigan.

**Supplementary Information**

**Supplementary Figures**

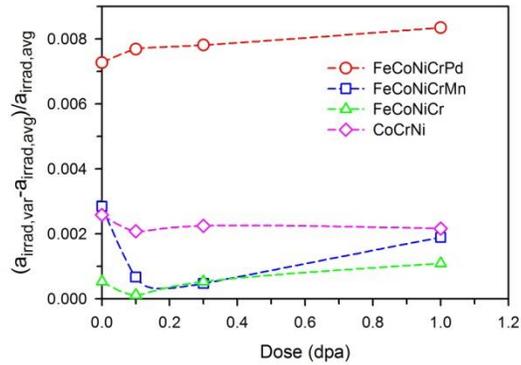

**Supplementary Figure 1. Evolution of local strain with dose.** Here, $a_{irrad,var}$ was extracted from the Rietveld refinement in a r range from 1.5 Å to 3 Å, covering only the 1st atomic shell.

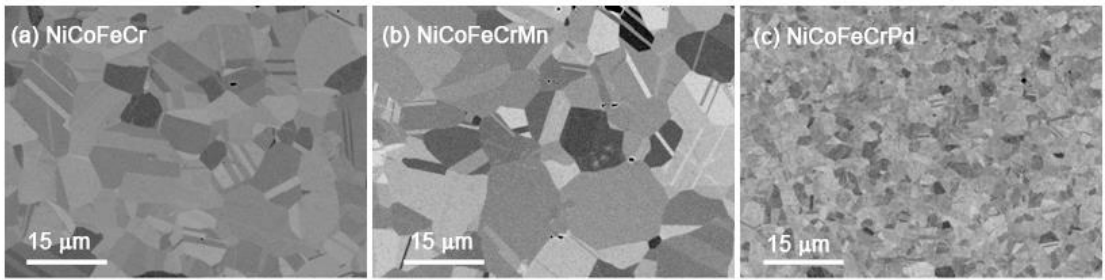

**Supplementary Figure 2. Microstructure of the prepared samples.** Backscattered electron images of (a) FeCoNiCr, (b) FeCoNiCrMn, and (c) FeCoNiCrPd.